\documentclass[12pt,preprint]{aastex}

\begin{document}


\title{STEREO/SECCHI Stereoscopic Observations Constraining the Initiation of Polar
Coronal Jets}
\author{S. Patsourakos\altaffilmark{1,2}}

\author{E. Pariat\altaffilmark{1,2,3}}

\author{A. Vourlidas\altaffilmark{1}}

\author{S. K. Antiochos\altaffilmark{3}}
\author{J. P. Wuelser\altaffilmark{4}}

\altaffiltext{1}{ Naval Research Laboratory, Space Science Division,
Washington,  DC 20375}
\altaffiltext{2}{Center for Earth Observing and Space Research, Institute for Computational 
Sciences - College of Science,
George Mason University, Fairfax, VA 22030}
\altaffiltext{3}{Space Weather Laboratory, NASA Goddard Space Flight Center, Greenbelt, MD 20774}
\altaffiltext{4}{Solar and Astrophysics Laboratory, Lockheed Martin ATC, 3251 Hanover Street, Palo Alto, CA 94304}



\begin{abstract}
  We report on the first stereoscopic observations of polar coronal
  jets made by the EUVI/SECCHI imagers on board the twin STEREO
  spacecraft.  The significantly separated viewpoints ($\sim$
  11$^\circ$ ) allowed us to infer the 3D dynamics and morphology of a
  well-defined EUV coronal jet for the first time.  Triangulations of
  the jet's location in simultaneous image pairs led to the true 3D
  position and thereby its kinematics. Initially the jet ascends
  slowly at $\approx$10-20 $\mathrm{{km}\,{s}^{-1}}$ and then, after
  an apparent 'jump' takes place, it accelerates impulsively to
  velocities exceeding 300 $\mathrm{{km}\,{s}^{-1}}$ with
  accelerations exceeding the solar gravity.  Helical structure is the most important
  geometrical feature of the jet which shows
  evidence of untwisting. The jet structure appears strikingly
  different from each of the two STEREO viewpoints: face-on in the one
  viewpoint and edge-on in the other. This provides conclusive evidence
  that the observed helical structure is real and is not resulting from
  possible projection effects of single viewpoint observations.  The
  clear demonstration of twisted structure in polar jets compares
  favorably with synthetic images from a recent MHD simulation of jets
  invoking magnetic untwisting as their driving mechanism.  Therefore,
  the latter can be considered as a viable mechanism for the
  initiation of polar jets.

\end{abstract}

 


\section{Introduction}

Polar coronal jets are collimated transient ejections of plasma occurring in polar coronal
holes. They were discovered in soft X-rays by SXT on \textsl{Yohkoh} (see Shibata et al.
(1996) for a review of the SXT results).
Recent  \textsl{Hinode\/} jet  observations revealed a higher occurrence frequency than previously believed (Cirtain et al. 2007). The 
new \textsl{Hinode\/} observations allowed to determine the statistical properties of jets (Savtseva et al. 2007),
and  to deduce their line of sight velocities and densities (Ciphor et al. 2007; 
Culhane et al. 2007 ; Kamio et al. 2007; Moreno Insertis, Galsgaard $\&$ Ugarte-Urra 2008).

Polar jets are believed to occur when reconnection between small scale, pre-existing or emerging, closed magnetic fields and the open large scale magnetic field of the 
coronal holes
take place, in the so-called {\it anemone-jets} model (e.g. the 2D simulations Yokoyama \& Shibata 1996). 
These numerical simulations, and the recent 3D one of  
Moreno Insertis et al. (2008)
were successful in reproducing 
several key observational aspects of polar jets such as their speeds and temperatures and their characteristic inverted $Y$ 
shape (e.g. Shibata et al. 1996). However, these simulations could not account for the important observation that a fraction of jets seems to exhibit 
helical structure and untwisting (e.g., Shimojo et al. 1996; Canfield et al. 1996 ; 
Pike $\&$ Mason 1998; Wang et al. 1998; Wilhelm, Dammasch $\&$ Hassler 2002; Jiang et al. 2007;
Filippov,  Golub $\&$ Koutchmy 2008).

A 3D model of polar jets has been recently proposed by Pariat, Antiochos $\&$ Devore (2008), (hereafter PAD08), with magnetic twist being the jet driver (see also Shibata $\&$ Uchida (1986) for a 2D twist model). Here we give only a brief description of the model
and its results.
In this simulation, a vertically oriented magnetic dipole, embedded below the photosphere, generates a null point in a corona which also contains a uniform-background vertical magnetic field. The resulting axisymmetric configuration contains
two distinct flux systems: a circular patch of strong closed magnetic flux surrounded by weaker open flux,  
leading to an axissymetric configuration.

Slow  horizontal motions, reaching 
a maximum of about 1 $\%$ of the local
Alfv\'en speed
at the line-tied photospheric feet of the closed magnetic  
flux  are applied, thereby twisting the closed field. 
The applied flow field conserves the distibution of the vertical
magnetic field $B_{z}$ and thus the axisymmetry, with an almost solid body rotation of the closed
flux. Initially, 
the  closed system responds quasi-statically to those  boundary motions by slowly growing vertically. 
Eventually a kink-like instability occurs which breaks the intitial symmetry. This triggers reconnection at the 
null-point and 
generates a massive jet with a speed at a fraction of the local Alfv\'en speed. 
The fact that reconnection remains dormant for a significant
ammount of time is an attractive element of the PAD08 model, since
this allows a significant energy build-up before the generation
of a massive and impulsive jet.

More importantly the model produces a 
non-linear  torsional wave
propagating outwards, which supplies
a continuous energy source for accelerating the jet to large
distances.
The wave is  due to the release of most of the twist
accumulated in the closed field region. 
Given that, in the corona,
the plasma is frozen-in to the magnetic field, the resulting torsional
wave could be traced by a helical structure undergoing untwisting
while the jet is lifting off. A significant amount ($\sim$ 90 $\%$) of
the helicity of the overall configuration is ejected with the
jet. 
Note finally that more recent simulations
with  non-axisymmetric configurations (with an inclined coronal field) lead
to quantitatively the same behavior (Pariat et al. 2008 in preparation).

However, we do not know whether helical structures seen in some jet
observations are 'real' or they result from projection effects introduced in 
single viewpoint observations. And what about
the true kinematic behavior of jets? Up to now, all jet velocity determinations
 were made from a single viewpoint thereby
determining a projected component on the plane of the sky only.
These important issues can only be addressed by using better
constraints on the geometry of the observed jets. STEREO observations,
with their two viewpoints, are thus perfectly suited to address these
questions. This Letter presents the first stereoscopic observations of polar
coronal jets. We determined the 3D kinematic evolution of 
a well-observed jet and studied its morphology
and more specifically  its
helical structure, how 
it differs in different vantage point 
and finally compared it with synthetic images from
the PAD08 model.

\section{Observations and Data Analysis}
We analyzed the observations of a polar jet
seen by the 2 STEREO spacecraft (hereafter A and B) over a northern 
polar coronal hole on June 7 2007 around 05:00 UT. The spacecraft separation 
was  $\approx 11.7^\circ$. We focus here on the images collected 
by the Extreme Ultraviolet Imaging Telescope
(EUVI) of the SECCHI (Wuelser et al. 2004; Howard et al. 2007)
suite of instruments (the jet was also
observed in the outer corona with the COR1 and COR2 coronagraphs).
EUVI takes full disk images in EUV channels
centered around 171, 195, 284 and 304 \AA \,
(hereafter  171, 195, 284 and 304).  
EUVI has $\sim$ 1.6 arcsec pixels and our observations have 
a cadence of 2.5, 10 and 20 minutes 
for  171, 195-304 and 284,  
respectively. The low-cadence of the 284 
data did not allow us to observe the generation of the jet.

We processed the images with the {\it secchi\_prep}
routine. Then, each synchronized image pair
\footnote{Since there are different light travel times
for A and B, EUVI image pairs are taken with a time difference
which  ensures they correspond to the {\it same} time on the Sun, i.e. they
are synchronized.} was co-aligned with the  routine {\it scc\_stereopair}.
This included scaling of both images to the same pixel size, shifting
to a common center and rolling to a common plane defined 
by the Sun center and the locations of the 2 spacecraft. 
Comparison of the limb locations (routine {\it euvi\_coalign})
and housekeeping data from the SECCHI Guide Telescope
lead to uncertainties in 
the  co-alignment of image pairs of  less than a pixel.

\begin{figure}[!ht]
\epsscale{0.5}
\plotone{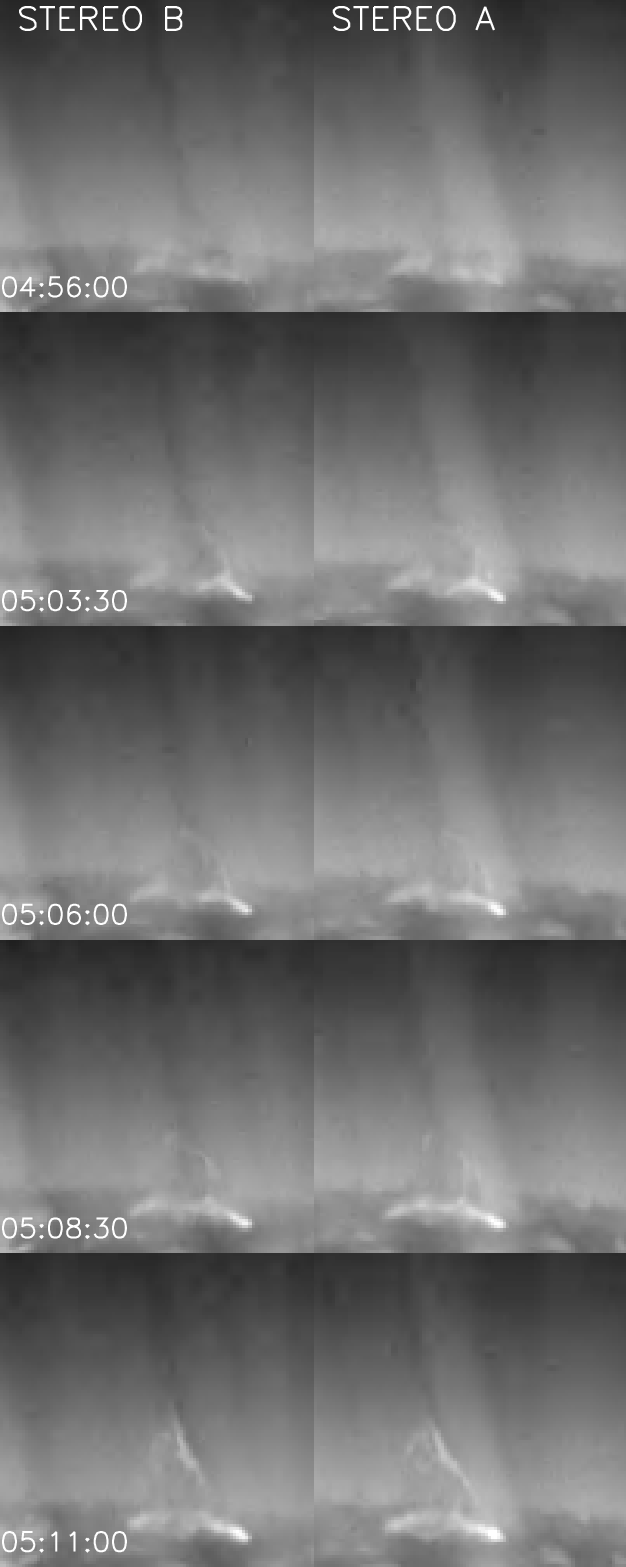}
\caption{Images
of the jet in the 171\AA\ channel for EUVI-A (left column) and
B (right column). All images have the same scaling and 
we plot the logarithm of the intensity which increases with color from black to white. 
Each image is $\approx$ 115 Mm on each side. Solar North is up.}
\label{fig:img}
\end{figure}

Figure \ref{fig:img} has synchronized image pairs of the jet 
in the 171  (see also the movies in the 171 and 195;
video1.mpg and video2.mpg). Figure \ref{fig:img} and the associated
movies show that the jet takes place in a configuration which includes
two bright points. The  western-most bright point starts to slowly
extend upwards at around 04:30, something that can be better seen in the 195. 
Then, at around 04:55 a brightening takes place 
at the base of this bright point and its structure starts to rise at a
faster rate. Sometime between 05:08 and 05:11 the rising structure appears to 'jump'
towards the left bright point and a jet, with the typical
inverted  $Y$ is formed. The base of the jet, corresponding to
the initial location of the two bright points, exhibits an intense brightening.

The jet develops an apparent helical structure in the last image 
pair of Figure \ref{fig:img} which appears similar on both spacecraft.
However, in the next, almost simultaneous 
image pairs in 195, 171 and 304, the situation is  markedly  different
(Figure \ref{fig:twist2}). The body of the jet seems to be viewed face-on in A 
and edge-on in B and appears to untwist while rising. Therefore, what
we observe could be a left-handed double helix for which
we see both threads in A whereas we see them apparently crossing in B.
These  significant  differences in the appearance of the jet between in A and B  
are  also visible in the next image pair in the 171, 
before the jet merges with the background, as it rapidly expands outwards
by seemingly untwisting. Both the helical jet body and its base appear very similar in
171 and 195, while small but noticeable differences can be seen in the
304. This implies that the jet contains both cool ($\approx$ 80000 K; 304) and
warm plasmas ($\approx$ 1-1.5 MK; 171 and 195).

Previous single-viewpoint inferences of helical structure in jets were
always susceptible to line of sight effects which could lead to false
conclusions. This is the first time that such a conclusive observation
on the existence of helical structures in jets is made. 
The wave-like motions we observed
could be related to  the transverse oscillations detected
in some jets (Cirtain et al. 2007), spicules (De                                                                                              
Pontieu  et al. 2007) 
and prominences (Okamoto et al. 2007).

Next we studied the 3D kinematics of the jet. We first visually
identified the tie-points of the jet-front (i.e. the image pixel
coordinates) in synchronized image pairs in 171 and 195, and repeated this
process 10 times for each considered image pair.  Triangulation of
those locations (routine {\it scc\_triangulate}; see Inhester (2006)
for the basics of triangulation and stereoscopy) supplied the 3D
coordinates (x,y,z) and the heliocentric distance of the jet front as
a function of time.  We then calculated the averaged radial distance
of the jet as a function of time $t$, $r_{jet}(t)$, and the
corresponding standard deviation (supplying an estimate of the error
bars) for the 10 tie-point selections for every considered image pair.

\begin{figure}[!h]
\epsscale{1.0}
\plotone{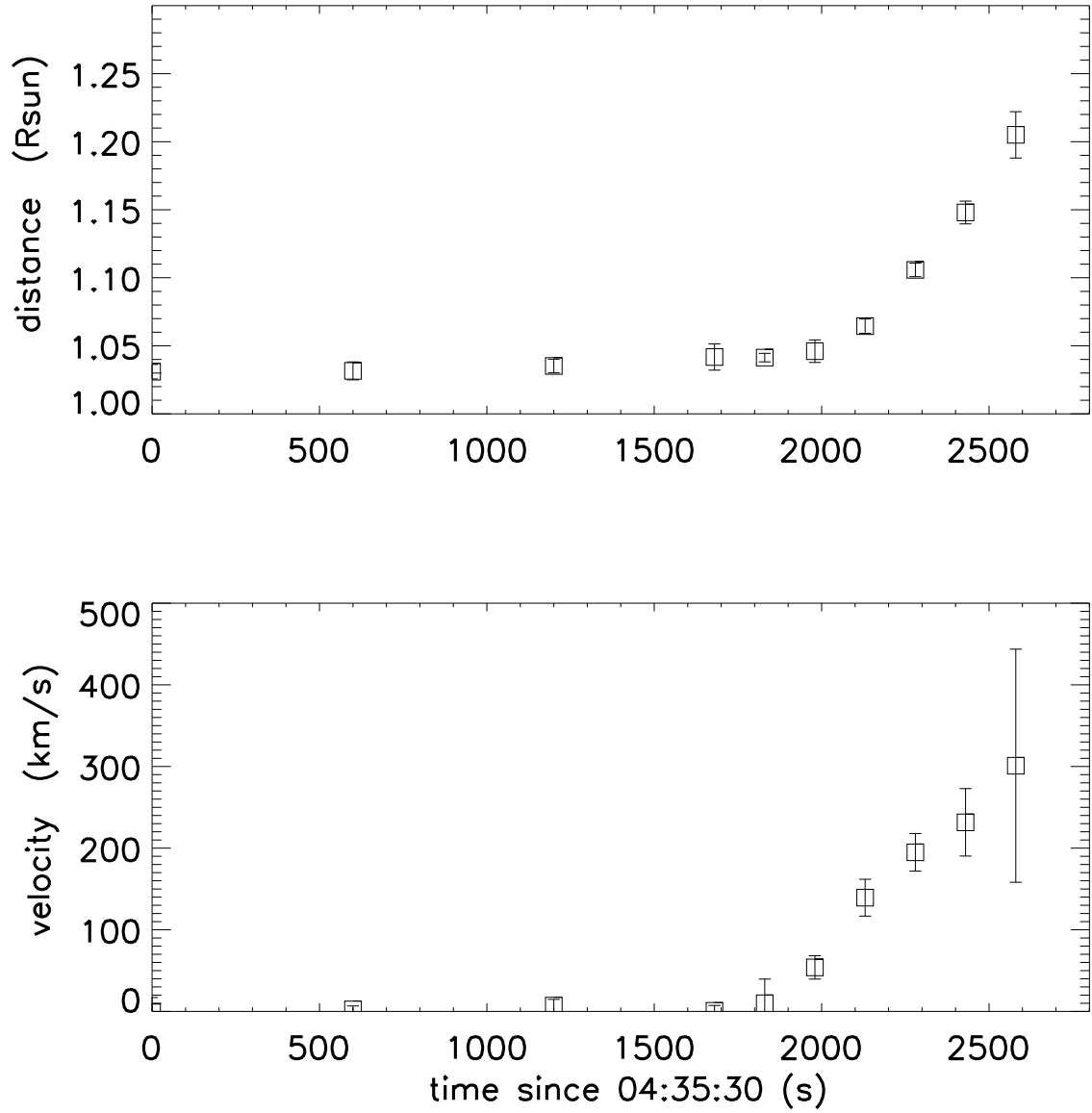}
\caption{Temporal evolution of the height and velocity of the jet
  front from stereoscopic measurements.} 
\label{fig:rv}
\end{figure}

Figure \ref{fig:rv} contains $r_{jet}(t)$ (upper panel) and the
temporal evolution of radial velocity of the jet (lower panel)
determined by numerical derivation of $r_{jet}(t)$ with respect to
$t$. This plot provides the 'true' distance and velocity of
the jet as opposed to determinations from a single viewpoint which
give access to a fraction of those quantities only. As can be seen in
Figure \ref{fig:rv} there are 2 phases in the kinematic evolution of
the jet: an initial phase ($0 \lesssim t \lesssim 2000 \, \mathrm{s}
$) during which the jet ascends very slowly at $\approx$ 10-20
$\mathrm{km\,{s}^{-1}}$ (note that for this phase we used data from
the 195 because of the better visibility of the jet in this line) followed by a
phase of impulsive acceleration ($ t \gtrsim 2000 \, \mathrm{s}
$) when the jet speeds up to  $\gtrsim$ 300
$\mathrm{km}\,{s}^{-1}$.  Such two-phase kinematic behavior is
characteristic of quasi-statically driven MHD systems which eventually
become instable.  The maximum speed of the jet is a substantial
fraction of the Alfv\'en speed in the corona.  The corresponding
accelerations, calculated from the second derivative of $r_{jet}(t)$
with respect to $t$, exceeded the solar gravity.

\begin{figure}[!h]
\epsscale{1.0}
\plotone{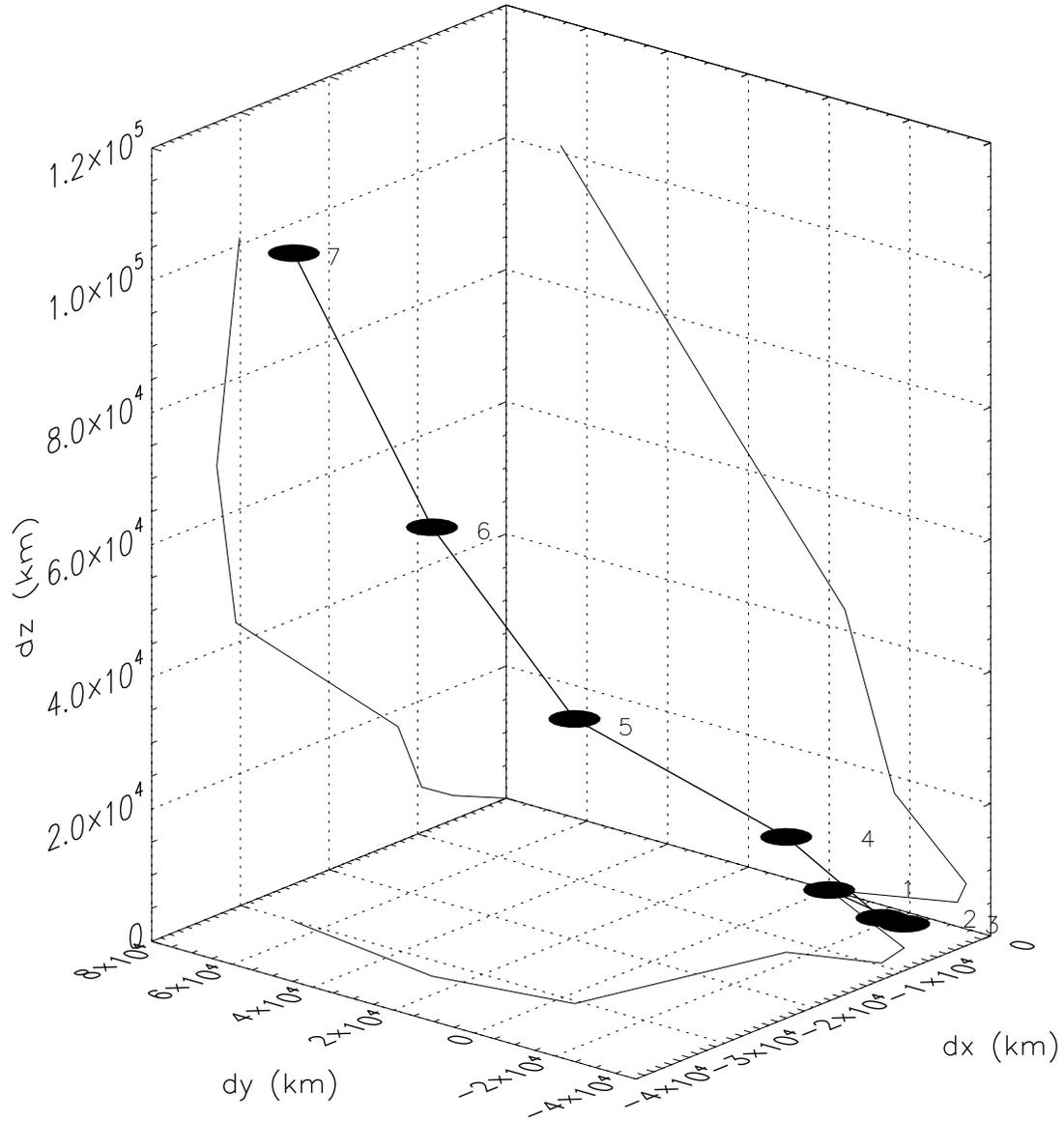}
\caption{3D trajectory (thick line connecting ellipses) of the jet
  during the impulsive acceleration phase (cf Figure \ref{fig:rv}).
  The x,y and z displacements of the jet with respect to the first
  point of the trajectory are plotted, (i.e. the trajectory starts at
  (0,0,0)). The projection of the orbit on the xy,yz, and xz planes is
  also shown (thin lines).}
\label{fig:3d}
\end{figure}

The deduced coordinates of the jet allowed us to trace its trajectory
in 3D. Figure \ref{fig:3d} has the jet trajectory
corresponding to the phase of rapid acceleration (i.e. the last 6
points of Figure \ref{fig:rv}).  The most striking feature is the
abrupt change in the jet plane from point 4 to point 5 of its orbit,
before following an almost vertical path in points 5 and 6.  This
change is correlated with the apparent 'jump' of the jet seen in the
images and the movies around 05:08 (4th row of Figure \ref{fig:img}),
thereby suggesting that a kink-like instability may have taken place.

\section{Discussion and Conclusions}
Our stereoscopic observations present the first conclusive evidence
of helical structures in polar jets. This lends significant support to
the PAD08 model, in which jets are driven by magnetic untwisting, and
predict the development of a clear helical structure.

\begin{figure}[!h]
\epsscale{1.1}
\plotone{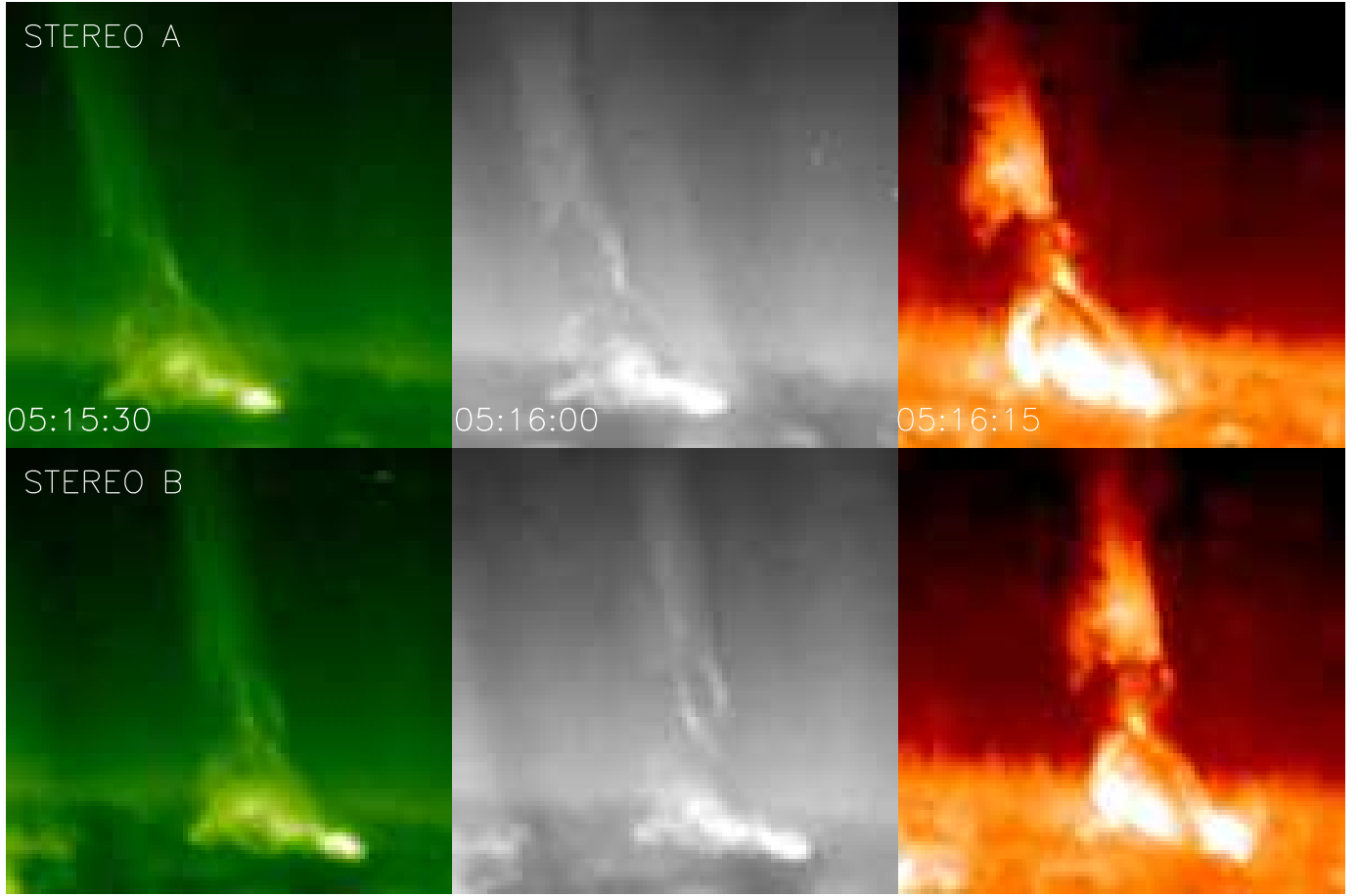}
\caption{
Simultaneous images 
of the jet in the EUVI 195\AA\, 171\AA\ and 304\AA\ channels (left, middle and right column, respectively) seen by STEREO-A and B (upper and lower row respectively).} 
\label{fig:twist2}
\end{figure}

\begin{figure}[!h]
\epsscale{1.1}
\plotone{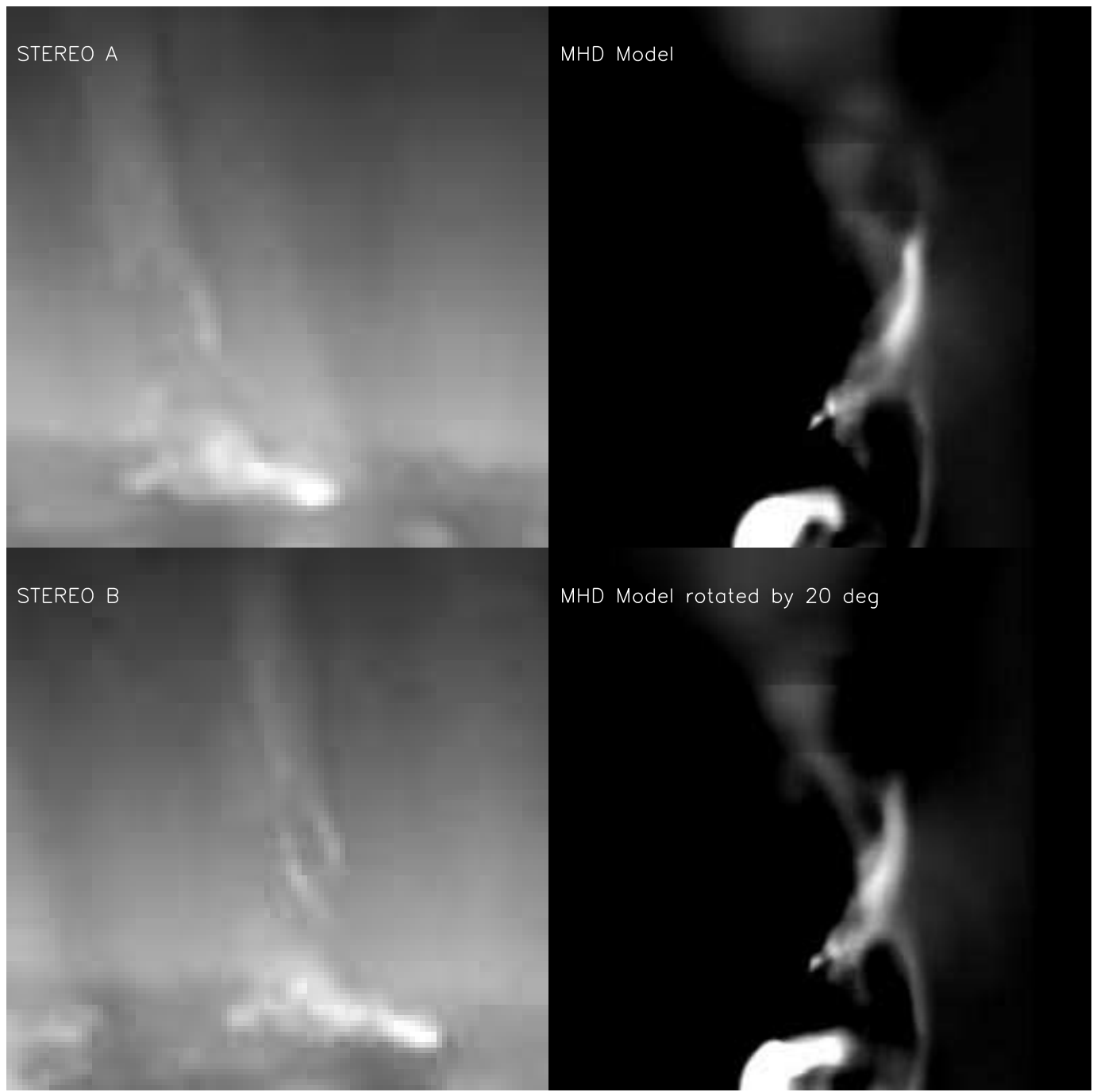}
\caption{Comparison of jet images in the 171 channel at 05:16
  UT seen by STEREO A and B (left column) with synthetic images from
  an MHD simulation of a jet viewed from 2 viewpoints separated by 20
  degrees (right column).}
\label{fig:twist1}
\end{figure}

For a more quantitative comparison between the helical jet structure
of our observations and the PAD08 simulations we produced synthetic
images from the latter. Results from a jet simulation having a
$10^\circ$ inclination of the volume coronal field (relatively to the
vertical; see PAD08) were used. We constructed synthetic images by
integrating the square of the electron density $n$ from the MHD
simulation, along horizontal lines of sight (orthogonal to the
direction of the jet).  This emulates the EUV emission that EUVI
records which is proportional to the line of sight integral of
$n^{2}$.  Temperature effects entering into the determination of EUV
intensities, through the temperature-dependent instrumental response
function, were not taken into account.  This assumption is likely to
be acceptable, because the instrumental temperature
response functions are rather narrow, tending to select quite
isothermal plasmas and we have shown that the jet has similar
morphology in 3 different EUVI channels (Figure \ref{fig:twist2}).
Synthetic images are  given in the right column of
Figure \ref{fig:twist1}.  The two simultaneous images correspond to
two viewpoints separated by 20 degrees, i.e.  the second image was
constructed by rotating the simulation box by 20 degrees about the
z-axis.  The images correspond to a snapshot of the PAD08 simulation,
shortly after the helical structure started to develop.  We note that
while in the top right image of Figure \ref{fig:twist1} two threads
can be seen, the bottom right image shows two crossing threads, very
similar to the face-on and edge-on views of the jet from A and B in
the left column of the same Figure respectively.

The field lines of the PAD08 model from the selected snapshot have
$\approx$ 1.1 turns or equivalently a twist of 2.2 $\pi$, a value that
should approximate the amount of twist present in the observed jet,
given the similarity between the synthetic and observed images in
Figure \ref{fig:twist1}.  This amount of twist is a lower
limit on the total twist applied to the observed configuration during
the initial slow evolution phase jet since there seems that
significant twist is left over at the base of the jet. The fact that a 
wider viewpoint separation has to be applied to the
simulation results in order to obtain a similar jet morphology with
respect to the observations is not surprising. The MHD simulations
starts with a relatively high degree of symmetry compared to what may
be expected for the 'real' Sun.

From the above comparison we conclude that magnetic twist is a viable
candidate for the initiation of polar jets. Moreover the kinematics of
the jet, with the slow rise phase, followed by a 'jump' and rapid
acceleration to a fraction of the Alfv\'en speed are also consistent
with the twist model of PAD08. However, similar kinematic behavior
results from models that do not invoke twist as their driver (e.g.,
Yokoyama \& Shibata 1996; Moreno Insertis et al. 2008).  It seems,
therefore, that the jet morphology can be the most sensitive
discriminator between different initiation mechanisms. We finally note 
that SXT observations showed that about 10$\%$ of jets
exhibit helical structure and untwisting (Shimojo et al. 1996).
Whether this fraction is a lower limit, due to the relatively
low spatial resolution of the SXT observations, should await detailed
statistical studies of the higher spatial resolution data from
\textsl{Hinode} and \textsl{SECCHI}.  

The SECCHI data used here were produced by an international consortium
of the Naval Research Laboratory (USA), Lockheed Martin Solar and
Astrophysics Lab (USA), NASA Goddard Space Flight Center (USA),
Rutherford Appleton Laboratory (UK), University of Birmingham (UK),
Max$-$Planck$-$Institut for Solar System Research (Germany), Centre
Spatiale de Li\`ege (Belgium), Institut d’ Optique Th\'eorique et
Applique\'e (France), and Institut d’Astrophysique Spatiale (France).
We thank the referee for useful comments.



\clearpage

\clearpage

\clearpage

\clearpage

\clearpage

\end{document}